# Optimal Control of Quantum Systems and a Generalized Level Set Method

**Fariel Shafee**
Department of Physics
Princeton University
Princeton , NJ 08544

**Abstract:**

We study the application of a generalized form of the level set method used in classical physical contexts to quantum optimal control situations. The set of OCT equations needed to keep the expectation value of an observable constant is first discussed and the dimensionality of the actual parameter space carefully considered. Then we see how concepts of level set methods emerge that may help solve the inverse problem associated with designing the control Hamiltonian with greater speed. The formal equations and the algorithm are presented.

# I. INTRODUCTION

We have previously reported [1-4] on the possible use of the level set method (LSM) [5-7] in a quantum context to keep the expectation value of an observable (EVO) constant. We have done so both to show how the LSM can be of relevance in keeping track of the EVO in state space, and also in the parameter space of the Hamiltonian. We have considered time as the motional parameter of the level sets, and also an arbitrary scale variable.

Here we report on the possible use of a method using level set concepts in an optimal control situation. In this case of optimal control, time appears naturally in the evolution of the Schroedinger equation. We introduce a set of scale variables $s_i$ to generalize the situation though it may be possible that in a particular physical problem only one such parameter is used. Such a scale parameter can be applied [8] to make a simple multiplicative change of the original variables of the Hamiltonian, or they may be used in a more complex fashion to design modulation of the variables [9]. In the next section we first see how OCT equations relate to a problem where a particular observable must be maintained with least possible departure from a set value. In the following section we see how equations like LSM equations evolve. In section 4 we present the algorithm briefly and in the last section we summarize our conclusions. Actual numerical work will be presented in a later report.

# II. OPTIMAL CONTROL OF AN EVO

The optimal control problem deals with finding the control Hamiltonian $H_{int}$ such that a cost function $C$ is minimized. The constituents of the cost function will be i) the departure from the set value of the observable at periods of measurement $t = n\,T$ beginning at t=0, ii) the cost of the control Hamiltonian, possibly measured in terms of the energy of the laser beam used iii) other possible considerations depending on the parameters of the problem.

To avoid losing transparency in a plethora of indices, wherever the context permits, we shall use the most concise possible notation to denote the sets of parameters in each class by a single letter. Thus the system Hamiltonian may be written as $H_o(a(s))$, or more briefly as $H_o(a)$ where s= $\{s_i\}$, $s_i$ being the set of scale parameters we have mentioned above, and the control Hamiltonian may be assigned the notation $H_c(b(s)) = H_c(b)$. If a parameter b appears in both $H_o$ and $H_c$
Then we shall identify it as a control parameter and consider the term in the system $H_o$ containing it as part of $H_c$.

In evolving from t=0 to t= T the system must satisfy Schroedinger equation (SE):

$$i\hbar \frac{\partial}{\partial t}|\Psi(t,a,b)> = H(a,b)|\Psi(t,a,b)> \quad \ldots (1)$$

where $a(s) \in R^n$ and $b(s) \in R^m$

in other words we have *n* system parameters and *m* control parameters, both of whom would depend on $s \in R^p$, i.e. on *p* scale parameters. The mappings

$$s \to a(s) \quad \ldots (2)$$

will be assumed as given by the physics of the system, but the mapping

$$s \to b(s) \quad \ldots (3)$$

is the design problem for our OCT, i.e. given the "scale" vector *s* in $R^p$, we know the system parameter vector *a* in $R^p$, but must then find the control vector *b* in $R^m$, such that the scalar cost function C is minimized.

Using the standard methodology of Lagrange multipliers for the SE constraint [10], we need to find variational functional equations from the constrained cost function C':

$$C'(a,b) = D(<\Psi(a,b,T)|\Theta|\Psi(a,b,T)>, \Theta_o) + \int dt\, I(E(b,t)) +$$

$$\int dt\, (<\lambda(a,b,t)|i\hbar\frac{\partial}{\partial t} - H(a,b)|\Psi(a,b,t)> + h.c.)$$

Where, as we have mentioned earlier,

$$H(a,b) = H_o(a) + H_c(b) \quad \ldots (4)$$

and in many cases the control is given by:

$$H_c(a,b) = \mu(a,b).E(b) \quad \ldots (5)$$

The deviation cost may be of the form in least squares:

$$D(a,b) = K\, (\Theta(T,a,b) - \Theta_o)^2 \quad \ldots (6)$$

with *K* a given constant, and $\Theta$ the preassigned value of the EVO.

The intensity cost of the control beam also may be of the square type, in many cases:

$$I(b,t) = L\, E(b,t)^2 \quad \ldots (7)$$

with L a suitable constant, *E* the electric field of the radiation. This function may be a quite complicated function of *b*, with variable frequencies and multiple pulses modulated as per an envelope dictated by the components of the control parameter vector b. Tactically it may be best to try a trial function for *E* guided by experience and the physical requirements of the problem concerned and then adjust that either parametrically or even functionally till a good fit is obtained, and then the OCT problem can find the exact values of the parameters $\{b_i\}$ for the job.

One can formally generalize further and consider a set of EVOs forming a vector in another space, and with a distance function for the whole vector from the set vector to form the argument of the relevant part of the cost function, as has been considered in ref.[10]. However, it is not necessary for us to introduce that complication in this work where we shall mostly concentrate on the techniques of the LSM.

The λ variation of course gives back the SE which needs to be satisfied at every time. And gives the prescription for the time evolution of the state function from t=0 to t=T, usually using numerical methods and/or approximation techniques.

The Ψ variation gives the equation for λ(t), which in this case agrees with the equation for the wave function Ψ, because of the cancellation of a potential extra term from <Theta> due to the presence of the h.c. part, though it might have been different if we had other parts in the cost function.

The final λ(T), however, is given by

$$i\hbar|\lambda(T,a,b)>=2K(\Theta(T,a,b)-\Theta_o)\hat{\Theta}|\Psi(T,a,b)>\quad\quad(8)$$

In principle we can begin with a trial state function, with trial values of b, calculate its value at t=T using SE, and thus get λ (T) from Eq. (8). Then we can use SE again to get λ(t) for 0<t<T. This may later be used to optimize the value of *b*.

## III. A GENERALIZED LEVEL SET METHOD

The gradient of the cost function with respect to the control vector b is needed to optimize b. like finding roots of an equation by Newton-Raphson method. In the present form this gradient is:

$$\nabla_b C' = 2L\int dt \mathbf{E}(b,t).\nabla_b \mathbf{E}(b,t)$$
$$+2\mu(a,b).\int dt\,\mathrm{Im}<\lambda(t,a,b)|\nabla_b \mathbf{E}(t,b)|\Psi(t,a,b)>\quad\quad(9)$$

It is likely that the dipole transition moment may not be a function of the control parameters, but in general it may be.

When the optimal value of C′ has been obtained the right hand side of Eq. (9) will be zero,

$$\nabla_b C' = 0 \quad\quad(10)$$

obviating any need for further changes of the last upgraded value of *b*. But in reality it may be necessary to obtain an assurance of a global minimum of the cost function by trying genetic algorithmic jumps.

At a fixed value of the scale vector *s*, the system parameters have fixed values *a(s)*, which we assume are known functions. So by the procedure given above we can find the corresponding b(s) vector in the control space. Equating the gradient to zero we actually get *m* equations in the m-dimensional control parameter space. Even if the equations are nonlinear we shall get a finite number of b vectors from the multiple roots, or, in the unlikely case of a transcendental equation a denumerably infinite number of root vectors. Except in trivial cases we shall have to obtain these roots numerically and hence approximately, but in many cases accuracy may be satisfactory in a reasonable number of steps.

On the other hand in principle we may have degenerate solutions. Let there be out of *m* equations let *d* be the number of independent equations. Then the optimal solution will correspond to a *(m-d)* dimensional surface. In particular in case of nonredundancy ( d=m), we shall have a set of discrete points, i.e. a set of zero measure. For one redundant equation we shall have a one dimensional curve etc. These are the level sets we are interested in and whose propagation with *s* we are interested in investigating.

In the LSM we try to determine the development of a level set curve (or a higher dimensional surface) as a whole, and not of individual points on the curve. This is ideally suited to our purpose in OCT, because we want to find a new optimal control vector b(s′) given an old value b(s), usually not bothering about which point on the curve we land in, though other additional constraints may make that a relevant concern in particular cases.

The optimal cost level set is therefore given implicitly by a number of equations related to the gradient of the cost function with respect to the control parameter vector for a particular set *s*. The dimensionality of the s vector does not seem to have any relevance to the dimensionality of the level sets in the *b* space. If the system parameters are functions of the different components of the s vector, a particular a-vector would correspond to an intersection of all these constant $a_i$ curves/surfaces in s-space, which means not all possible a-vectors may be allowed, but the inverse is not true; given any s-vector we can find an a-vector and hence the corresponding b-vectors[s], which will be a discrete set, as discussed above.

Since it is somewhat improbable that different gradient equations Eq. 10 will not be independent, except in trivial cases, let us explore another possibility of having continuous level sets for the optimal control parameters. If the system paramters also include some parameters not dependent on the scale vector *s* (let us call them the set *c*) then the solutions to Eq. 10 will be of the form

*b* = *b (s, c)* ……………………………………………………………………………….(11)

So, for the same s value we shall get the optimal control parameter for different values of the components of the *c*-vector at different points of the b-space, and the dimensionality of this set will be the dimensionality of *c*. it is necessary to remind ourselves that the points on this level set do not correspond to the same value of the cost function. But each *b*-vector on this set does correspond to an optimal value of the control parameter vector *b* for some vector *c* of the system parameters that do not depend on the scale vector *s*.

This is a quite plausible situation, though it will be necessary to discover by some means the values of the scale independent parameters *c* in the original system Hamiltonian.

As usual for LSM, the propagation of the level sets with the scale variable *s* can be parametrized by a speed normal to the level set curve or surface. The standard procedure comprises the following steps:
1) to find level sets for different *s* values at discrete a values giving discrete b sets which need to be smoothly joined by some interpolating method [8,11-13] in *b*-space;
2) to find the normals to the level sets;
3) to find the normal "speeds" of the curves/surfaces from the b-vectors at different discrete s values;
4) To interpolate, or less reliably, to extrapolate to arbitrary s values.

The advantage of the level set method is that we do not need to go through the expensive procedure of recalculating the *b* values for any *s*. An initial investment on a reasonably sized discrete arrray can hopefully give us the control parameters for any *s*. This can be a distinct advantage in real-time experiments.

### III.    THE ALGORITHM

Before we can do some actual calculation( to be reported later) let us collect together the sequential steps of the procedure for this method of optimal control to keep the EVO reasonably constant subject to constraints.
1) We should know how the system Hamiltonian $H_o$ depends on the parameters a, and how the parameters a depend on the scale parameter(s) *s*. We should also know how *H* depends on unscaled parameters *c*.
2) We should begin with an initial trial functional form of the field **E** in terms of time and the control parameter vector *b*, which may control many different features of the field, e.g. duration, frequency variation, amplitude, amplitude variation etc.
3) We can now calculate the first trial wave function using SE from a trial initial value at *t=0* to one at time *T*.
4) We can now find the value of the Lagrange multiplier lambda at time *T*.
5) We can then find by inverting SE lambda for any time *0<t<T*.
6) We can find using Eq. 9 the gradient of the cost function in terms of the b-vector.
7) We can use this gradient to refine the initial guess of the *b* vector
8) Repeat steps 3 to 7 with the refined b until there is no significant change.
9) Repeat 3 to 8 for other values of the unscaled parameters *c*.
10) Repeat 3 to 9 for another value of s.
11) Find the level sets at discrete s values.
12) Find the normals at different points of the curves and the "speeds".
13) Store and use this information to get the control *b* values for any *s* and *c*.

# IV. CONCLUSIONS

Even though the LSM may not be the best, or sometimes probably even a relevant, method in OCT, it does provide an interesting alternative that deserves full investigation. The steps involved in the LSM part are not more expensive than in most other methods, and may indeed be quite fast, as in the case of fluid dynamics calculations or epitaxial crystal growth prediction. In a future report we shall try to show explicit calculations to systems where it is applicable, with comparison to other methods.

I would like to thank Professor H. Rabitz for interesting discussions.

# REFERENCES

1. F. Shafee, "Quantum control with the Level Set Method" (unpublished, 2002)
2. F. Shafee, "Energy level sets for the Morse potential" (unpublished, 2002)
3. F. Shafee, "Evolution of level sets in Hamiltonian parameter space" (unpublished, 2003)
4. F. Shafee, "Quantum optimal control and level sets" (unpublished, 2002)
5. J. Sethian, "The evolution of level set and fast marching methods", http://math.berkeley.edu/~sethian
6. H.H. Thodberg, "The level set method", http://www.imm.dtu.dk/courses/02503/Lectures
7. C.H. Phillips, "The level set method", http://web.mit.edu/aram/www/work/thesis.pdf
8. J.M. Geremia and H. Rabitz, " Global nonlinear algorithm for inverting quantum mechanical observations", *Phys. Rev.* **A 64**, 022710-1-13 (2001)
   J.S. Biteen, J.M. Geremia and H. Rabitz, " Optimal quantum control field design using logarithmic maps", *Chem. Phys. Lett.* **483**, 440-446 (2001)
9. A. Mitra and H. Rabitz, "Identifying mechanisms in the control of quantum dynamics through Hamiltonian encoding", Princeton preprint (2002)
10. J. Botina, H. Rabitz, and N. Rahman, "A simplified approach to optimally controlled quantum dynamics*", J. Chem. Phys.* **104**, 4031-4040 (1996)
11. H. Bartels, J.C. Beatty and B.A. Barsky, *An Introduction to Splines for Use in Computer Graphics and Geometric Modelling.* (Morgan Kaufmann, San Francisco, CA, 1998).
12. G. Micula and S. Micula, *Handbook of Splines.* (Kluwer, Dordrecht, Netherlands, 1999).